\documentclass[10pt,conference]{ieeeconf}

\IEEEoverridecommandlockouts 

\usepackage[utf8]{inputenc}
\usepackage{textcomp}

\usepackage{graphicx}
\usepackage{amsmath,mathtools}
\usepackage{amssymb}
\usepackage{pifont}
\usepackage[version=4]{mhchem}
\usepackage{siunitx}
\usepackage{longtable,tabularx}
\usepackage[dvipsnames]{xcolor}
\usepackage{pgfplots}
\usepackage{pgfplotstable}
\usepgfplotslibrary{fillbetween}
\usepgfplotslibrary{groupplots}
\pgfplotsset{compat=1.18}
\usepackage{tikz}
\usepackage[caption=false,font=normalsize,labelfont=sf,textfont=sf]{subfig}
\usetikzlibrary{shapes.geometric, positioning}

\newtheorem{remark}{Remark}

\newtheorem{proposition}{Proposition}
\newtheorem{theorem}{Theorem}

\newcommand{\mc}[0]{\mathcal}
\renewcommand{\bf}[0]{\mathbf}
\newcommand{\argmax}[0]{\operatorname*{arg\,max}}
\newcommand{\im}[0]{\operatorname{Im}}
\newcommand{\bing}[2]{\operatorname{Bing}_{#1,#2}}

\title{Bayesian Learning of Uncertain Linear System Behaviours}

\author{
  Noreen Abdelwahab, %
  Alberto Padoan, %
  and Dominic Liao-McPherson%
  \thanks{The authors are with the University of British Columbia, Vancouver, BC, V6T 1Z4, Canada (emails: 
          {\tt\small noreen.abdelwahab@ubc.ca, alberto.padoan@ubc.ca, dliaomcp@mech.ubc.ca}). This work was supported by the Swiss Federal Office of Energy (SI/502734) and the Natural Sciences and Engineering Research Council of Canada (ALLRP 591401 - 23) through the MAESTRO project. }%
}

\begin{document}
\maketitle

\begin{abstract}
	We present a Bayesian approach to learning uncertain behavioural representations of linear dynamical systems from noisy input-output data. In behavioural systems theory, the finite-horizon behaviour of a linear system is represented by a subspace. We quantify uncertainty in this subspace through probability distributions on the Grassmann manifold. This leads to a geometric formulation of Bayesian inference over the Grassmannian. We derive the posterior distribution under Gaussian measurement noise and discuss how uncertainty in the inferred behaviour can be propagated to trajectory predictions, with a view towards data-driven control.
\end{abstract}

\section{Introduction}
Behavioural systems theory~\cite{willems_time_1986} characterizes a linear time-invariant (LTI) system through its subspace of admissible trajectories. A representation of this subspace can be constructed from sufficiently informative input--output data~\cite{willems_note_2005}, without first identifying a parametric model. Over finite horizons, the restricted behaviour is represented by the column space of a suitably constructed data matrix. This perspective supports methods for system identification~\cite{van_overschee_subspace_1996,markovsky_algorithms_2005,verhaegen_filtering_2007} and data-enabled control~\cite{coulson_data-enabled_2019,coulson_distributionally_2022,markovsky_behavioral_2021}, with demonstrated practical effectiveness~\cite{markovsky_behavioral_2021,markovsky_data-driven_2023}.

In practice, noise corrupts both the offline training data and the online measurements used for feedback and prediction, and can severely degrade controller performance if its effects are not accounted for~\cite{dorfler_bridging_2023}. Quantifying and compensating for these effects is therefore a fundamental challenge in both system identification and data-driven control. This has motivated formulations that extend beyond the deterministic, noise-free setting~\cite{markovsky_behavioral_2021}, including regularized optimization~\cite{coulson_data-enabled_2019,dorfler_bridging_2023}, robust formulations~\cite{huang_robust_2023}, and distributionally robust formulations~\cite{coulson_distributionally_2022}, which can improve robustness but often introduce hyperparameters that require tuning. Other strategies preprocess data before constructing the behavioural representation, for example through low-rank approximation~\cite{markovsky_low_2012}, including truncated singular value decomposition (SVD)~\cite{markovsky_data-driven_2022}. Maximum a posteriori estimation offers another approach to trajectory inference under uncertainty~\cite{yin_unified_2026}.

We propose an uncertainty quantification framework for restricted behaviours of linear dynamical systems. The central idea is to represent uncertainty about the restricted behaviour by a probability distribution on the Grassmann manifold and learn this distribution from data using Bayesian inference. Motivated by~\cite{padoan_behavioral_2022,padoan_distances_2025},  this formulation respects the invariance of the subspace to the choice of basis and quantifies uncertainty arising from noisy observations. It also allows uncertainty to be propagated to finite-horizon trajectory predictions using tools from geometric probability and computation on manifolds~\cite{girolami_riemann_2011,tran_variational_2021,lin_bayesian_2017,turaga_statistical_2008}.

The statistical analysis of estimation error and model uncertainty has long been central to system identification~\cite{ljung_system_1999}. Uncertainty quantification for identified LTI models has been developed for state-space~\cite{chiuso_numerical_2004,tsiamis_statistical_2023,mesquita_robust_2025}, frequency-response~\cite{damjanovic_distributionally_2026,pedari_error_2026}, and ARMA representations~\cite{soderstrom_system_1989}. For predictors constructed from behavioural representations, uncertainty in the predicted outputs is quantified in~\cite{yin_data-driven_2022,kaviani_uncertainty_2025}, and a deterministic analysis of prediction error is given in~\cite{jin_sensitivity_2026}. Related work has also investigated uncertainty quantification and geometric distances for restricted behaviours~\cite{padoan_behavioral_2022,padoan_distances_2025}.

Closely related research provides complementary perspectives on probabilistic descriptions of dynamical systems. Faulwasser et al.~\cite{faulwasser_behavioral_2023} use polynomial chaos expansions (PCE) to obtain a deterministic representation of stochastic dynamics in a higher-dimensional coefficient space. Sasfi et al.~\cite{sasfi_gaussian_2025} model finite-length trajectories as Gaussian random vectors, characterizing trajectory variability through their covariance structure. Finally, Preciado~\cite{preciado_measure-theoretic_2026} develops a measure-theoretic formulation based on probability measures supported on admissible trajectories, leading to convex formulations of optimal control problems. Our focus is Bayesian inference over restricted behaviours themselves, with uncertainty in the inferred behaviour propagated to trajectory predictions.

\section{Preliminaries}\label{prelim}
Let $\mathbb{S}^q_{\ge 0}$ and $\mathbb{S}^q_{>0}$ be the sets of
$q\times q$ symmetric positive semidefinite and positive definite
matrices, respectively. For a matrix $A$, we write $A^\top$ for its
transpose, $A^\dagger$ for its Moore--Penrose pseudoinverse, $|A|$
for its determinant, and $\sigma_i(A)$ for its $i$th largest singular value. The $k\times k$ identity matrix is $I_k$.
We use the abbreviations
\[
\operatorname{etr}(A):=\exp(\operatorname{tr}(A)),
\qquad
\|x\|_M^2:=x^\top Mx,
\]
where $M$ is symmetric positive definite. The Gaussian distribution
with mean $\mu\in\mathbb{R}^n$ and covariance
$\Sigma\in\mathbb{S}^n_{>0}$ is written $\mathcal{N}(\mu,\Sigma)$,
with density
\[
\mathcal{N}(y;\mu,\Sigma)
:=(2\pi)^{-n/2}|\Sigma|^{-1/2}
\exp\!\left(-\tfrac12\|y-\mu\|_{\Sigma^{-1}}^2\right).
\]

We will make use of two types of smooth manifolds in this paper: the compact Stiefel manifold and the Grassmann manifold~\cite{absil_riemannian_2004}  (see also~\cite{boumal_introduction_2023,absil_optimization_2008}). Subspaces are denoted by bold uppercase letters and their orthonormal basis matrices by the corresponding uppercase letters. For integers $n>0$ and $k>0$,  with $k\leq n$, the Grassmann manifold \(\operatorname{Gr}(n,k)\) is the set of all $k$-dimensional linear subspaces of \(\mathbb{R}^n\)
\[
\operatorname{Gr}(n,k)
=
\left\{
\mathbf{U}\subseteq\mathbb{R}^n:
\dim(\mathbf{U})=k
\right\} .
\]
The compact Stiefel manifold $\operatorname{St}(n,k)$ is the set of all matrices in $\mathbb{R}^{n\times k}$ with orthonormal columns
\[
\operatorname{St}(n,k)
=
\left\{
U \in\mathbb{R}^{n\times k}:
U^\top U=I_k
\right\}.
\]
The compact Stiefel manifold is an embedded submanifold of
$\mathbb{R}^{n\times k}$, and the Grassmann manifold is its quotient
under orthogonal changes of basis $\operatorname{Gr}(n,k)
\cong \operatorname{St}(n,k)/\mathcal{O}(k)$~\cite{absil_riemannian_2004,absil_optimization_2008},
where $\mathcal{O}(k)$ is the orthogonal group of $k\times k$
matrices. Thus, a subspace ${\mathbf U\in\operatorname{Gr}(n,k)}$
is represented by any ${U\in\operatorname{St}(n,k)}$ whose columns
span $\mathbf U$. Two representatives $U$ and $V$ describe the
same subspace if and only if $V=UQ$ for some $Q\in\mathcal{O}(k)$.

We can also identify $\mathbf U\in\operatorname{Gr}(n,k)$ with its
orthogonal projector $P_{\mathbf U}=UU^\top$, which is independent
of the representative $U$~\cite{chikuse_statistics_2003,bendokat_grassmann_2024}.
For $\mathbf V=\operatorname{im}V$ with $V\in\operatorname{St}(n,k)$,
the chordal distance is defined by
\[
d_{\mathrm c}(\mathbf U,\mathbf V)
:=\frac{1}{\sqrt2}\|P_{\mathbf U}-P_{\mathbf V}\|_F,
\]
where $\|\cdot\|_F$ is the Frobenius norm. Equivalently,
\[
d_{\mathrm c}(\mathbf U,\mathbf V)^2
=\sum_{i=1}^k\sin^2\theta_i,
\]
where ${\theta_i\in[0,\tfrac{\pi}{2}]}$, with
${\theta_1\leq\cdots\leq\theta_k}$, are the principal angles between
the subspaces $\mathbf U $ and $\mathbf V $, and satisfy
$\cos\theta_i=\sigma_i(U^\top V)$~\cite{conway_packing_1996,edelman_geometry_1998}.

The compact Grassmann manifold $\operatorname{Gr}(n,k)$ carries a uniform probability measure $\nu$~\cite{chikuse_statistics_2003}. Throughout the paper, densities on $\operatorname{Gr}(n,k)$ are taken with respect to $\nu$. For a symmetric matrix
$A\in\mathbb R^{n\times n}$, the matrix Bingham distribution
on $\operatorname{Gr}(n,k)$ has density~\cite{chikuse_statistics_2003}
\begin{equation}\label{eq:bingham-def}
  \operatorname{Bing}_{n,k}(\mathbf U;A)
  :=\frac{\operatorname{etr}(AP_{\mathbf U})}
  {{}_1F_1\bigl(\tfrac k2;\tfrac n2;A\bigr)},
\end{equation}
where ${}_1F_1$ is the confluent hypergeometric function of
matrix argument~\cite{muirhead_aspects_2005}, which can be
evaluated numerically using the algorithms
in~\cite{koev_efficient_2006}.

\section{Problem Setup}\label{probSetup}
We consider a discrete-time LTI system
\begin{subequations}\label{eq:SS}
	\begin{align}
		x_{t+1} &= Ax_t + Bu_t, \label{eq:part_a}\\
		y_t &= Cx_t + Du_t, \label{eq:part_b}
	\end{align}
\end{subequations}
where $x_t\in\mathbb{R}^n$, $u_t\in\mathbb{R}^m$, and $y_t\in\mathbb{R}^p$ denote the state, input, and output, respectively. The system matrices $A$, $B$, $C$, and $D$ are unknown, and the state is unobserved. 

For integers $t\ge1$ and $L\ge1$, define the stacked input
and output vectors
\[
\begin{aligned}
u_{[t,L]} &:=
\begin{bmatrix}
u_t^\top & \cdots & u_{t+L-1}^\top
\end{bmatrix}^\top,\\
y_{[t,L]} &:=
\begin{bmatrix}
y_t^\top & \cdots & y_{t+L-1}^\top
\end{bmatrix}^\top.
\end{aligned}
\]
The stacked vector
\[
w_{[t,L]} :=
\begin{bmatrix}
u_{[t,L]}\\
y_{[t,L]}
\end{bmatrix}
\]
is an admissible length-$L$ input--output trajectory
of~\eqref{eq:SS} if there exists a state sequence
$x_t,\ldots,x_{t+L}$ satisfying~\eqref{eq:SS} at each time
$\tau=t,\ldots,t+L-1$. The restricted behaviour
$\mathcal B_L$ of~\eqref{eq:SS} consists of all admissible
input--output trajectories $w_{[1,L]}$ of~\eqref{eq:SS}. The restricted behaviour
$\mathcal B_L$ of~\eqref{eq:SS} consists of all admissible
input--output trajectories $w_{[1,L]}$ of~\eqref{eq:SS}. The restricted behaviour $\mc B_L$ is a linear subspace of $\mathbb R^d$, where $q=m+p$ and $d=qL$. We denote its dimension by $r$. For an observable realization~\eqref{eq:SS}, $r=mL+n$ for $L\ge\ell$, where $\ell$ is the observability index of $(C,A)$~\cite{willems_note_2005}.

Our objective is to infer the restricted behaviour of~\eqref{eq:SS} from noisy observations and provide a principled quantification of uncertainty in the inferred subspace. Inputs and outputs are assumed to be measured with additive Gaussian noise, giving an errors-in-variables model~\cite{soderstrom_errors--variables_2018}. Specifically, we consider the probabilistic model
\begin{equation}
  w_i=\bar w_i+\varepsilon_i,
\qquad
\bar w_i\in\mc B_L,
\qquad i=1,\ldots,N.
\end{equation}
where $\varepsilon_i\in\mathbb{R}^d$ is independent of the input-output trajectory $\bar w_i$ and such that\footnote{For noise covariance $\sigma_\varepsilon^2\Sigma$ with known $\Sigma\succ0$, the transformation $w\mapsto\Sigma^{-1/2}w$ makes the noise isotropic and maps $\mc B_L$ to $\Sigma^{-1/2}\mc B_L$.}
$\varepsilon_i\overset{\mathrm{i.i.d.}}{\sim}\mathcal N(0,\sigma_\varepsilon^2I_d)$.

\subsection{Subspace Representation from Data}
\label{deterministic}
We first recall how a restricted behaviour can be represented in the absence of noise. 
Consider a measured trajectory
\[
w_{\mathrm d}(t) =
\begin{bmatrix}
	u_{\mathrm d}(t)\\
	y_{\mathrm d}(t)
\end{bmatrix}
\in\mathbb{R}^{q},
\qquad t=1,\ldots,T .
\]
For a positive integer $L\leq T$, define the depth-$L$ Hankel matrix
$\mathcal{H}_L(w_{\mathrm d})\in\mathbb{R}^{qL\times (T-L+1)}$, by
\[
\mathcal{H}_L(w_{\mathrm d}) =
\begin{bmatrix}
	w_{\mathrm d}(1) & w_{\mathrm d}(2) & \cdots
	& w_{\mathrm d}(T-L+1)\\
	w_{\mathrm d}(2) & w_{\mathrm d}(3) & \cdots
	& w_{\mathrm d}(T-L+2)\\
	\vdots & \vdots & \ddots & \vdots\\
	w_{\mathrm d}(L) & w_{\mathrm d}(L+1) & \cdots
	& w_{\mathrm d}(T)
\end{bmatrix}.
\]
If $(A,B)$ is controllable and $u_{\mathrm d}$ is persistently exciting of order $L+n$, the fundamental lemma~\cite{willems_note_2005,van_waarde_willems_2020} gives
\begin{equation}\label{eq:span_condition}
    \mc B_L=\operatorname{im}\mathcal{H}_L.
\end{equation}
Thus, every admissible trajectory $w\in\mc B_L$ can be written as
\[
w=\mathcal{H}_Lg
\]
for some $g\in\mathbb R^{T-L+1}$. The construction extends to multiple records by concatenating their Hankel matrices under a collective excitation condition~\cite{van_waarde_willems_2020}. For length-$L$ input--output trajectories, the corresponding noise-free data matrix represents $\mc B_L$ whenever it has rank $r$.

An orthonormal representation follows from the compact SVD~\cite{markovsky_behavioral_2021},
\[
\mathcal{H}_L=U_r\Sigma_rV_r^\top,
\]
where $U_r\in\mathbb R^{d\times r}$ satisfies $U_r^\top U_r=I_r,
$ and $
\mc B_L=\operatorname{im}U_r.$
Thus, the restricted behaviour is a point on $\operatorname{Gr}(d,r)$~\cite{padoan_behavioral_2022,padoan_distances_2025}, and represented by any of its orthonormal basis matrices. For noisy measurements, the rank of a Hankel matrix may be larger than the dimension of underlying noise-free restricted behaviour. In this case, standard practice is to truncate the singular value decomposition by retaining $r$ dominant singular directions to obtain an estimate of the subspace~\cite{markovsky_behavioral_2021}.

\section{Probabilistic Beliefs over Behaviours}\label{methods}
The core idea in this paper is to model finite-horizon behaviours as elements of the Grassmann manifold and to learn the belief 
\begin{equation}
	p(\mc{B}_L~|~W),
\end{equation}
where  $W \in\mathbb{R}^{d\times N}$ is a data matrix of observed trajectories
\begin{equation} \label{eq:data_matrix}
W=\begin{bmatrix}w_1&\cdots&w_N\end{bmatrix}
\in\mathbb{R}^{d\times N}.
\end{equation} 
Uncertainty about the restricted behaviour $\mc B_L$ is modelled by a probability distribution\footnote{We equip the Grassmann manifold with its Borel $\sigma$-algebra and take densities with respect to the uniform probability measure $\nu$.} on $\operatorname{Gr}(d,r)$.
Thus, our framework replaces a deterministic restricted behaviour with a probability measure over admissible behaviours. A distribution can subsequently be propagated into probabilistic behavioural predictive control and simulation problems, providing a principled mechanism for representing epistemic uncertainty in data-driven predictions.

\subsection{Inference} \label{inference}
Let $\mathbf S\in\operatorname{Gr}(d,r)$ denote the unknown restricted behaviour and let $\Gamma\in\operatorname{St}(d,r)$ be any representative, so that $\mathbf S=\operatorname{im}\Gamma$ and $P_{\mathbf S}=\Gamma\Gamma^\top$. We use the observation model
\begin{equation}\label{eq:observation_model}
	w_i = \Gamma z_i + \varepsilon_i, \qquad i=1,\ldots,N,
\end{equation}
where all $z_i, \varepsilon_i$ are independent, $z_i\sim\mathcal N(0,\sigma_z^2I_r)$ and $\varepsilon_i\sim\mathcal N(0,\sigma_\varepsilon^2I_d)$, with $\sigma_z^2>0$ being a hyperparameter (usually estimated using expectation maximization) and $\sigma_\varepsilon^2>0$ being a known noise covariance. This model \eqref{eq:observation_model} generates noisy samples
\begin{equation}
  w_i\sim\mathcal N(0,\sigma_z^2P_{\mathbf S}+\sigma_\varepsilon^2I_d)
\end{equation}
from $\mathbf S$. Noise-free samples $w_i = \Gamma z_i \sim \mathcal N(0,\sigma_z^2P_{\mathbf S})$ satisfy $w_i \in \mathbf{S}$ with a distribution that does not depend on the specific representative $\Gamma$ but only the rotation-invariant projection $P_{\mathbf S}$. The latent vector $z_i$ can be thought of as a prior on the initial state of the system and serves as a hyperparameter. The choice of $\mu_z\in\mathbb{R}^d$ is not critical (as a prior its influence will be forgotten as the dataset grows), so we take $\mu_z = 0$ throughout for simplicity. The variance of $z_i$ must be isotropic however, if instead $z_i\sim\mathcal N(0,\Lambda)$ with a fixed $\Lambda \not \propto I$, then the noise-free trajectories depend on the representation $\Gamma$ rather than the subspace $\mathbf{S}$.

By Bayes' rule, the posterior density of $\mathbf S$ with respect to $\nu$ is
\begin{equation}\label{eq:bayes}
	p(\mathbf S\mid W)=\frac{p(W\mid\mathbf S)\,p(\mathbf S)}{p(W)},
\end{equation}
with 
$ p(W)=\int_{\operatorname{Gr}(d,r)}p(W\mid\mathbf T)\,p(\mathbf T)\,d\nu(\mathbf T). $
Next, we examine the likelihood $p(W\mid \Gamma)$, prior $p(\Gamma)$, and normalizing constant (or evidence) $p(W)$.

\subsubsection{The likelihood} We first introduce the sample covariance matrix and aggregated precision, defined as
\begin{equation} \label{eq:Sw}
S_W:=\frac{1}{N-1}WW^\top,\quad \alpha:=\frac{N-1}{2}\,\frac{\sigma_z^2}{\sigma_\varepsilon^2(\sigma_\varepsilon^2+\sigma_z^2)},
\end{equation}
respectively. The next result characterizes the likelihood.

\begin{proposition}[Likelihood]\label{prop:likelihood}
Consider the observation model \eqref{eq:observation_model} and the data matrix $W$ defined in~\eqref{eq:data_matrix}. Assume that $\sigma_z^2>0$ and $\sigma_\varepsilon^2>0$ are known. Then, $p(W\mid \mathbf{S}) \propto \operatorname{Bing}_{d,r}(\alpha S_W)$, i.e., for every $\mathbf S\in\operatorname{Gr}(d,r)$,
\begin{align}
p(W \mid \mathbf S) &= C_1 \operatorname{etr}(\alpha S_W P_{\mathbf S}) \label{eq:l-bingham}
\end{align}
where $S_W$ and $\alpha$ are defined in \eqref{eq:Sw} and 
\begin{align*}
C_1 &= \left( (2\pi)^d (\sigma_\varepsilon^2)^{d-r} (\sigma_\varepsilon^2+\sigma_z^2)^r \right)^{-\frac{N}{2}} \exp\Bigl(-\frac{N-1}{2\sigma_\varepsilon^2} \operatorname{tr}(S_W)\Bigr).
\end{align*}
\end{proposition}

\begin{proof}
	Fix a representative $\Gamma$ of $\mathbf S$. By \eqref{eq:observation_model}, the $w_i$ are i.i.d.\ with $w_i\sim\mathcal N(0,\Sigma_w)$, $\Sigma_w=\sigma_z^2P_{\mathbf S}+\sigma_\varepsilon^2I_d$. Since $P_{\mathbf S}$ is an orthogonal projector of rank $r$, $|\Sigma_w|=(\sigma_\varepsilon^2+\sigma_z^2)^r(\sigma_\varepsilon^2)^{d-r}$ and $\Sigma_w^{-1}=\sigma_\varepsilon^{-2}I_d-\frac{2\alpha}{N-1}P_{\mathbf S}$~\cite{tipping_probabilistic_1999}. Hence
	\[
	p(W\mid\mathbf S)=(2\pi)^{-\frac{Nd}{2}}|\Sigma_w|^{-\frac N2}\operatorname{etr}\Bigl(-\frac{N-1}{2}\Sigma_w^{-1}S_W\Bigr),
	\]
	which is \eqref{eq:l-bingham}, and depends on $\Gamma$ only through $P_{\mathbf S}$.
\end{proof}

Proposition~\ref{prop:likelihood} shows that the log-likelihood is an affine function of $P_{\mathbf S}$ and that the data enter only through $S_W$ and, as a function of $\mathbf S$, the likelihood is proportional to a matrix Bingham density. The likelihood can also be naturally expressed using projectors. Some algebraic manipulation yields
\[ 
p(W\mid\mathbf S)= C_1 \operatorname{etr}(\alpha S_W)  \exp \left(-\frac{\alpha}{N-1} M(\mathbf S)\right),
\] where $M(\mathbf S) = \sum_{i=1}^N\|(I_d-P_{\mathbf S})w_i\|^2$. $M(\mathbf S)$ is known as the ``misfit''  and the terms $\|(I - P_{\mathbf S})w_i\|$ are a measure of distance between the data $w_i$ and the subspace $\mathbf S$ \cite{markovsky_exact_2006}.

\begin{remark}(Maximum Likelihood Estimation) The Maximum Likelihood Estimate (MLE) 
\begin{equation}
  \hat{\mathbf{S}}_{MLE} \in \argmax_{\mathbf S\in\operatorname{Gr}(d,r)} p(W\mid \mathbf{S})
\end{equation}
provides some insights about the likelihood. By Ky Fan's maximum principle~\cite{fan_theorem_1949} $\hat{\mathbf{S}}_{MLE} = \im~\hat \Gamma_r$, where $\hat \Gamma_r$ is a matrix of the $r$ dominant Eigvenvectors of $S_W$. This equivalent to the PCA estimate of~\cite{tipping_probabilistic_1999} and the column space of the truncated SVD of the data matrix used in subspace identification~\cite{van_overschee_subspace_1996} and in subspace predictive control~\cite{favoreel_spc_1999}. The subspace $\hat{\mathbf{S}}$ is unique if and only if $\lambda_r>\lambda_{r+1}$, where $\lambda_1\ge\cdots\ge\lambda_d$ are the eigenvalues of $S_W$. This condition holds almost surely under Gaussian noise whenever $N > r$, i.e., we have enough data relative to the rank of the system.
\end{remark}

\subsubsection{Selection of the prior}
We use priors of the form
\begin{equation}
  p(\bf S) = \bing{d}{r}(\bf S; A_0),
\end{equation}
in the matrix Bingham family \eqref{eq:bingham-def}. We consider two specific choices of the prior. The choice $A_0=0$ gives the uninformative uniform prior $p(\mathbf S_0)=1$, that is, the invariant distribution $\nu$. Alternatively, let $\mathbf S_0\in\operatorname{Gr}(d,r)$ be a reference subspace with projector $P_0 = \Gamma_0 \Gamma_0^\top$ obtained for example by a SVD of an independent data matrix $W_{\mathrm d}$. Then the informative prior
\begin{equation}\label{eqn:prior}
	p(\mathbf S)= \bing{d}{r}(\bf S; \kappa P_0) = e^{\kappa r}\exp\bigl(-\kappa\,d_{\mathrm c}(\mathbf S,\mathbf S_0)^2\bigr)
\end{equation}
is a decreasing function of the chordal distance to $\mathbf S_0$ and has mode $\mathbf S_0$. The hyperparameter $\kappa \geq 0$ thus quantifies trust in $\bf S_0$ and \eqref{eqn:prior} reduces to the uniform prior for $\kappa=0$. 

\subsubsection{The evidence}
For both choices of prior, the evidence $p(W)$ in \eqref{eq:bayes} cannot be written in closed-form but can be expressed in terms of hypergeometric functions. With the uniform prior, 
$$p(W)=C_1\,{}_1F_1(\frac r2;\frac d2;\alpha S_W)$$ 
by Proposition~\ref{prop:likelihood} and \eqref{eq:bingham-def}. With the prior \eqref{eqn:prior}, 
$$p(W)=C_1\,{}_1F_1(\frac r2;\frac d2;\alpha S_W+\kappa P_0)/{}_1F_1(\frac r2;\frac d2;\kappa P_0).$$ 
Both functions can be evaluated with the algorithm of~\cite{koev_efficient_2006} or by Monte Carlo integration with exact samples from $\nu$~\cite{mezzadri_how_2007}.

\subsubsection{The posterior distribution}
With the likelihood and prior in hand, we can derive expressions for the posterior.

\begin{theorem}[Posterior]\label{thm:posterior}
Consider the observation model \eqref{eq:observation_model}, the data matrix $W$, and $S_W$ and $\alpha$ as in \eqref{eq:Sw}. Assume that $\sigma_z^2>0$ and $\sigma_\varepsilon^2>0$ are known and that the prior is $\operatorname{Bing}_{d,r}(A_0)$ for a symmetric $A_0\in\mathbb R^{d\times d}$. Then the posterior is distributed according to
\begin{equation}\label{eq:postBingham}
p(\mathbf S\mid W) \sim \operatorname{Bing}_{d,r}(A_0+\alpha S_W)
\end{equation}
In particular, the uniform prior gives $\operatorname{Bing}_{d,r}(\alpha S_W)$ and the prior \eqref{eqn:prior} gives $\operatorname{Bing}_{d,r}(\alpha S_W+\kappa P_0)$.
\end{theorem}

\begin{proof}
By Proposition~\ref{prop:likelihood} and \eqref{eq:bingham-def}, $p(W\mid\mathbf S)\,p(\mathbf S)$ equals $\operatorname{etr}\bigl((A_0+\alpha S_W)P_{\mathbf S}\bigr)$ times a factor that does not depend on $\mathbf S$. Normalizing with respect to $\nu$ by \eqref{eq:bingham-def} gives \eqref{eq:postBingham}.
\end{proof}

Theorem~\ref{thm:posterior} shows that the matrix Bingham family is conjugate to the likelihood of Proposition~\ref{prop:likelihood}. Conjugacy follows because the log-likelihood is affine in $P_{\mathbf S}$, the sufficient statistic of the exponential family $\bing{d}{r}$ on $\operatorname{Gr}(d,r)$. The data enter through the additive update $A=A_0+\alpha S_W$. Since $\alpha$ grows linearly with $N$, a fixed prior contribution $A_0$ becomes negligible relative to the data term as $N\to\infty$.

\begin{remark}(The MAP Estimate)
The maximum a posteriori (MAP) estimate is
\[
\hat{\mathbf S}_{\mathrm{MAP}}
\in
\operatorname*{arg\,max}_{\mathbf S\in\operatorname{Gr}(d,r)}
p(\mathbf S\mid W).
\]
By Ky Fan's principle, it can be computed as
$\hat{\mathbf S}_{\mathrm{MAP}}=\operatorname{im}V_r$,
where the columns of $V_r$ are eigenvectors associated with the $r$ largest eigenvalues of $A$. For the prior $A_0=\kappa P_0$, the MAP estimate coincides with the MLE when $\kappa=0$ and converges to $\mathbf S_0$ as $\kappa\to\infty$. The concentration of the posterior around the MAP estimate is governed by the gaps between the $r$ largest eigenvalues of $A$ and the remaining eigenvalues.

The condition~\eqref{eq:span_condition} for noise-free data has a natural statistical counterpart. One may require the distribution of the underlying trajectories to span the restricted behaviour, i.e., 
\begin{equation}\label{eq:trajectory-informativeness}
\operatorname{im}\Sigma_{\bar w}=\mc B_L,
\qquad
\Sigma_{\bar w}:=\mathbb E[\bar w_i\bar w_i^\top].
\end{equation}
Under i.i.d. sampling with finite second moments and isotropic measurement noise, the MLE then converges almost surely to $\mc B_L$ as $N\to\infty$. The MAP estimate has the same limit for a fixed Bingham prior. The observation model~\eqref{eq:observation_model} satisfies this condition automatically.
\end{remark}

\subsection{Prediction}\label{prediction}
Let $T_{\mathrm{ini}}+T_{\mathrm f}=L$ and let 
\begin{equation}
  w=\Gamma z+\varepsilon
\end{equation}
be a new trajectory, where $(z,\varepsilon)$ is independent of $W$ and distributed as \eqref{eq:observation_model}. We permute and partition the trajectory as $w=(w_a,y_{\mathrm f})$, where $w_a=(u_{\mathrm{ini}},y_{\mathrm{ini}},u_{\mathrm f})\in\mathbb R^{d_a}$ collects the initial inputs $u_{\mathrm{ini}}\in\mathbb R^{mT_{\mathrm{ini}}}$, the initial outputs $y_{\mathrm{ini}}\in\mathbb R^{pT_{\mathrm{ini}}}$ and the future inputs $u_{\mathrm f}\in\mathbb R^{mT_{\mathrm f}}$, and $y_{\mathrm f}\in\mathbb R^{pT_{\mathrm f}}$ collects the future outputs we wish to predict. We also partition a representative and the projector,
\[
\Gamma=\begin{bmatrix}\Gamma_1\\ \Gamma_2\end{bmatrix},\qquad
P_{\mathbf S}=\begin{bmatrix}P_{11}&P_{12}\\ P_{21}&P_{22}\end{bmatrix}=\begin{bmatrix}\Gamma_1\Gamma_1^\top&\Gamma_1\Gamma_2^\top\\ \Gamma_2\Gamma_1^\top&\Gamma_2\Gamma_2^\top\end{bmatrix},
\]
in the same way. The following theorem characterizes the predictive distribution.

\begin{theorem}[Predictive distribution]\label{thm:predict}
	Consider the observation model \eqref{eq:observation_model}, the data matrix $W$, and the new trajectory $w=(w_a,y_{\mathrm f})$. Assume that $\sigma_z^2>0$ and $\sigma_\varepsilon^2>0$ are known and let $\mathbf S\mid W \sim \bing{d}{r}(A)$ be our belief about $\bf S$. Then:\smallskip
    
	\noindent i) For any fixed $\bf S$, $(y_f \mid w_a, \bf{S}) \sim \mc{N}(m(\bf S), \Sigma(\bf S))$ with
    \begin{align*}
  m(\mathbf S)&=\sigma_z^2P_{21}\bigl(\sigma_z^2P_{11}+\sigma_\varepsilon^2I\bigr)^{-1}w_a,\\
  \Sigma(\mathbf S)&=\sigma_z^2P_{22}+\sigma_\varepsilon^2I-\sigma_z^4P_{21}\bigl(\sigma_z^2P_{11}+\sigma_\varepsilon^2I\bigr)^{-1}P_{12},
\end{align*}

\noindent ii) the predictive density is
		\begin{multline}
        p(y_{\mathrm f}\mid w_a,W) =\\ \int_{\operatorname{Gr}(d,r)}\mathcal N\bigl(y_{\mathrm f};m(\mathbf S),\Sigma(\mathbf S)\bigr)\,p(\mathbf S\mid w_a,W)\,d\nu(\mathbf S), \label{eq:predictive}
		\end{multline}
		where\footnote{Replacing $p(\mathbf S\mid W,w_a)$ by $p(\mathbf S\mid W)$ in~\eqref{eq:predictive} omits the additional information about $\mathbf S$ supplied by $w_a$. The approximation is accurate when $p(w_a\mid\mathbf S,W)$ is approximately constant over the region where $p(\mathbf S\mid W)$ concentrates, as may occur for sufficiently large $N$.} $p(\mathbf S\mid w_a,W)\propto\mathcal N\bigl(w_a;0,\sigma_z^2P_{11}+\sigma_\varepsilon^2I\bigr)\,p(\bf S\mid W)$.
\end{theorem}

\begin{proof}
	i) By the proof of Proposition~\ref{prop:likelihood}, $w\mid\mathbf S\sim\mathcal N(0,\sigma_z^2P_{\mathbf S}+\sigma_\varepsilon^2I_d)$, and $m(\mathbf S)$ and $\Sigma(\mathbf S)$ are the conditional mean and covariance of a partitioned Gaussian vector. Since $\|P_{21}\|\le1$ and $\sigma_z^2P_{11}+\sigma_\varepsilon^2I\succeq\sigma_\varepsilon^2I$, the bound on $m(\mathbf S)$ holds. Writing $y_{\mathrm f}=\Gamma_2z+\varepsilon_{\mathrm f}$, where $\varepsilon_{\mathrm f}$ is independent of $(z,w_a)$, gives $\Sigma(\mathbf S)=\operatorname{Cov}(\Gamma_2z\mid w_a,\mathbf S)+\sigma_\varepsilon^2I\succeq\sigma_\varepsilon^2I$.
	ii) Since $w$ and $W$ are conditionally independent given $\mathbf S$, $p(y_{\mathrm f}\mid w_a,\mathbf S,W)=p(y_{\mathrm f}\mid w_a,\mathbf S)$ and $p(\mathbf S\mid w_a,W)\propto p(w_a\mid\mathbf S)\,p(\mathbf S\mid W)$. The claim follows from the law of total probability and Theorem~\ref{thm:posterior}.
\end{proof}

Theorem~\ref{thm:predict} expresses the predictive density as a continuous mixture of Gaussian densities weighted by the posterior $p(\mathbf S\mid W,w_a)$. The resulting density need not be Gaussian and may be multimodal when substantial uncertainty about the restricted behaviour remains. The integral over the manifold can be evaluated by quadrature in low dimensions or approximated using Riemannian Monte Carlo methods in higher dimensions.

When the posterior is concentrated near a point estimate $\hat{\mathbf S}$, such as the MLE or MAP estimate, one may use the approximation
$$
p(y_{\mathrm f}\mid w_a,W)
\approx p(y_{\mathrm f}\mid w_a,\hat{\mathbf S}),
$$
where the right-hand side is the Gaussian density characterized in Theorem~\ref{thm:predict}.
For fixed $\mathbf S$, the noise-free limit $\sigma_\varepsilon\to0$ gives
$$
\begin{aligned}
m(\mathbf S)
&\to \Gamma_2\Gamma_1^\dagger w_a
=P_{21}P_{11}^\dagger w_a,\\
\Sigma(\mathbf S)
&\to\bar\Sigma
=\sigma_z^2\Gamma_2
(I_r-\Gamma_1^\dagger\Gamma_1)\Gamma_2^\top.
\end{aligned}
$$
If $\Gamma_1$ has full column rank, then $\bar\Sigma=0$ and, for compatible $w_a\in\operatorname{im}\Gamma_1$, the limiting prediction is
$$
y_{\mathrm f}=\Gamma_2\Gamma_1^\dagger w_a.
$$
This recovers the noise-free predictor of~\cite{markovsky_data-driven_2008,favoreel_spc_1999}, evaluated at $\mathbf S$. For $\mathbf S=\mc B_L$, the full column rank condition holds if $T_{\mathrm{ini}}\geq l(\mathbf S)$, where $l(\mathbf S)$ is the system lag.

\section{Numerical Case Study}\label{num}
Consider a first-order autonomous system 
\begin{align*}
    x_{t+1} &= ax_t\\
    y_t &= x_t,
\end{align*}
where \(a=0.8\), with additive measurement noise on the output according to \(\varepsilon\sim \mathcal{N}(0, \sigma_\varepsilon^2)\) and \(\sigma_\varepsilon=0.15\). We choose this system because its restricted behaviour for $L=2$ is a line in $\mathbb R^2$, allowing direct visualization of uncertainty in the inferred subspace. We generate \(N\) trajectories of length \(L=2\). We rewrite the system as
\[w=\begin{bmatrix}
    y_0 \\ y_1
\end{bmatrix}
=\begin{bmatrix}
    x_0 \\ 
    a x_0
\end{bmatrix}+\varepsilon=\Gamma z + \varepsilon \]

A basis for the noise-free trajectory subspace is deduced from this representation as 
\[\Gamma_\mathrm{true} = \frac{1}{\sqrt{1+a^2}}
\begin{bmatrix}
1\\a
\end{bmatrix} =
\begin{bmatrix}
\cos\theta \\
\sin\theta 
\end{bmatrix}\]
yielding the noise-free subspace as a line through the origin angled at \(\theta_0\approx 38.66^\circ\).

\subsubsection{Inference} 
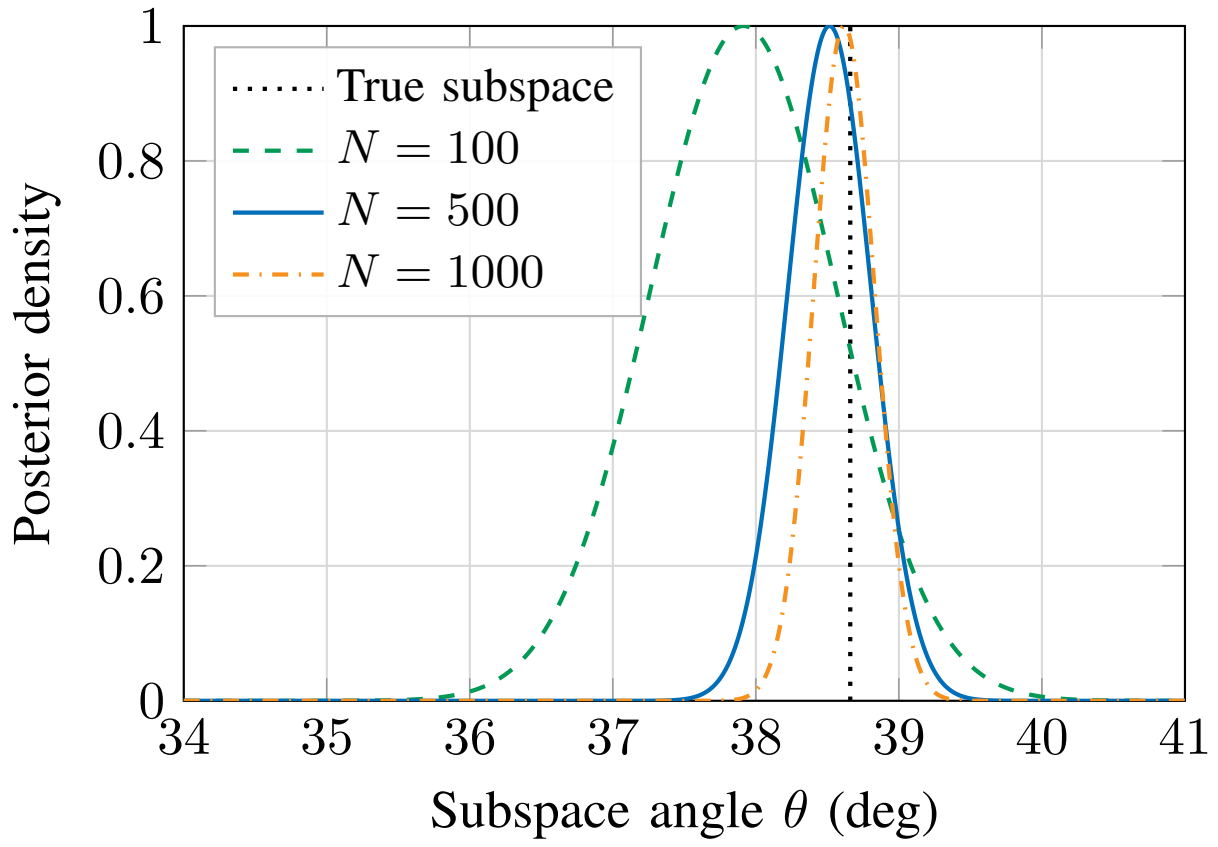
\begin{figure}[htbp]
  \centering
  \begin{tikzpicture}
    \begin{axis}[
      width=0.95\columnwidth,
      height=0.7\columnwidth,
      xmin=34, xmax=41,
      ymin=0,    ymax=1,
      xlabel={Subspace angle $\theta$ (deg)},
      ylabel={Posterior density},
      grid=major,
      grid style={gray!30},
      axis background/.style={fill=white},
      legend pos=north west,
      legend cell align={left},
      legend style={
        draw=gray!60,
        fill=white,
        fill opacity=0.9,
        text opacity=1,
        font=\small
      }
    ]
      \addplot[color=black, dotted, thick] coordinates {(38.6598, 0) (38.6598, 1)};
      \addlegendentry{True subspace}
      \addplot[color=ForestGreen, dashed, thick, smooth] table[header=false, col sep=comma, x index=0, y index=1] {post.txt};
      \addlegendentry{$N=100$}
      \addplot[color=NavyBlue, thick, smooth] table[header=false, col sep=comma, x index=0, y index=2] {post.txt};
      \addlegendentry{$N=500$}
      \addplot[color=BurntOrange, dash dot, thick, smooth] table[header=false, col sep=comma, x index=0, y index=3] {post.txt};
      \addlegendentry{$N=1000$}
    \end{axis}
  \end{tikzpicture}
  \caption{Posterior density with true subspace $\theta_0 = 38.66^\circ$ and MAP estimates $\hat{\theta}= 37.92^\circ$, $38.51^\circ$ and $38.62^\circ$ for $N=100$, $500$ and $1000$ samples, respectively.}
  \label{fig:posterior_distribution}
\end{figure}

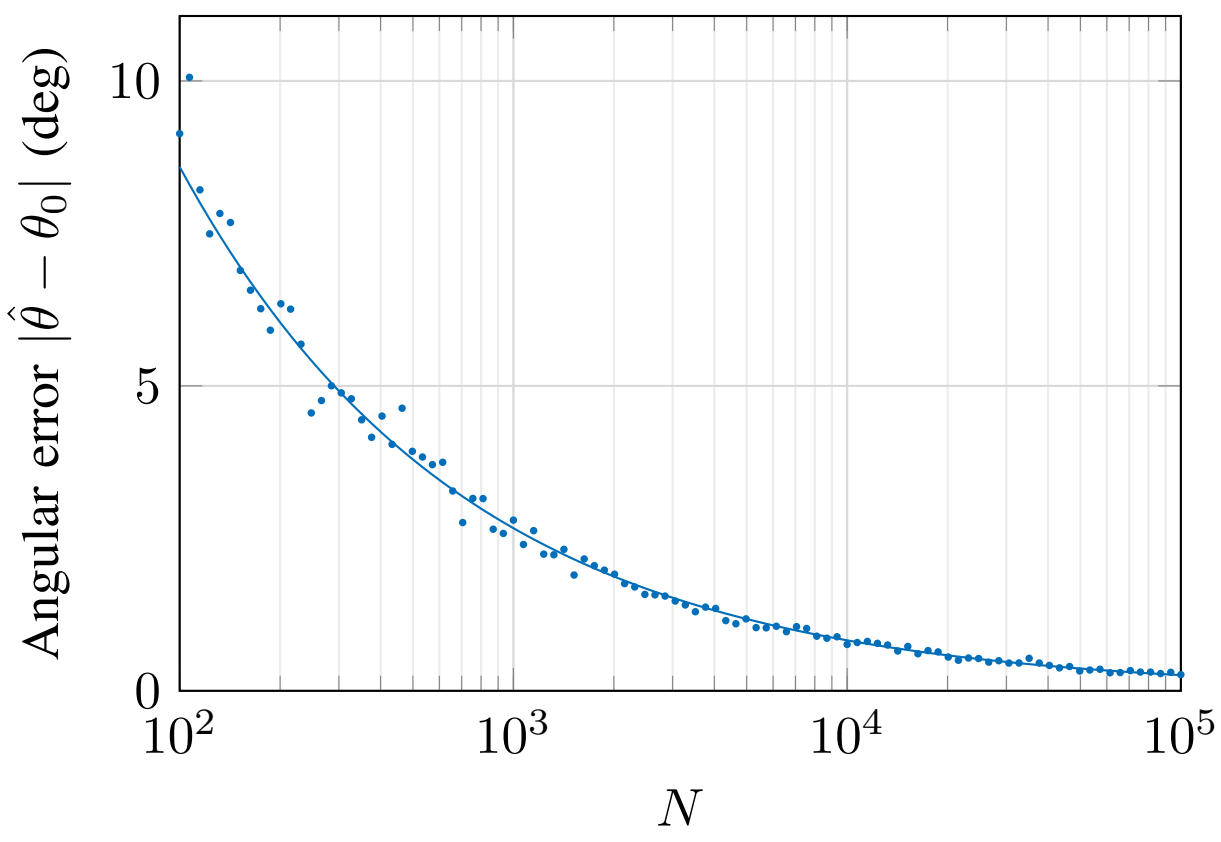
\begin{figure}
    \centering
\begin{tikzpicture}
\begin{axis}[
    width=0.95\columnwidth,
    height=0.7\columnwidth,
    xmode=log,
    xmin=10^2, xmax=10^5,
    ymin=0,
    xlabel={$N$},
    ylabel={Angular error $|\hat{\theta} - \theta_0|$ (deg)},
    grid = both,
    major grid style={gray!30},
    minor grid style ={gray!15},
      legend pos=north east,
      legend cell align={left},
      legend style={
        draw=gray!60,
        fill=white,
        fill opacity=0.9,
        text opacity=1,
        font=\small
      }
]

    \addplot[color=NavyBlue, only marks, mark=*, mark size=0.5pt] table[
        header=false, col sep=comma, x index=0, y index=1] {angular_err_vs_N_2.txt};
    \addplot[NavyBlue, samples=100, domain=10^2:10^5]{89.135*x^(-0.508)};

\end{axis}
\end{tikzpicture}
\caption{Principal angle error between MAP subspace estimate and true subspace as a function of $N$ for $\sigma_\varepsilon=0.15$ with corresponding trendline. The error decreases as $N^{-1/2}$. Each data point represents the average over multiple independent noise realizations to obtain a more representative estimate of the underlying trend.}
    \label{fig:angular_error_vs_N}
\end{figure}

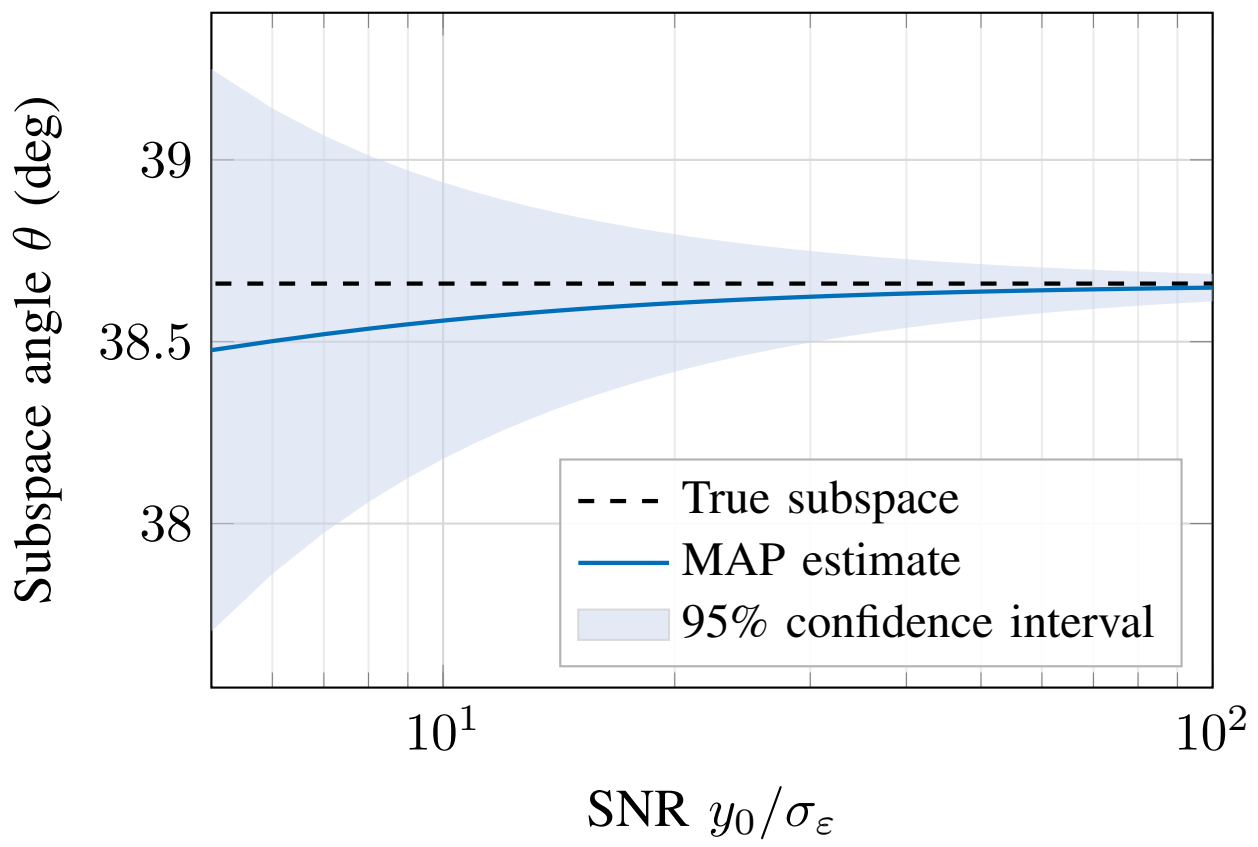
\begin{figure}[htbp]
    \centering
    \begin{tikzpicture}
    \begin{axis}[
    width=0.95\columnwidth,
    height=0.7\columnwidth,
    xmode=log,
    xmin=5, xmax=100,
    xlabel={SNR $y_0/\sigma_\varepsilon$},
    ylabel={Subspace angle $\theta$ (deg)},
    grid = both,
    major grid style={gray!30},
    minor grid style ={gray!15},
      legend pos=south east,
      legend cell align={left},
      legend style={
        draw=gray!60,
        fill=white,
        fill opacity=0.9,
        text opacity=1,
        font=\small
      }
]
             \addplot[black, dashed, thick, samples=2, domain=1:100] {38.6598}; 
            \addlegendentry{True subspace}  

            \addplot[name path=upper1, draw=none, forget plot] 
                table[header=false, col sep=comma, x index=0, y index=2] {data.txt};
            
            \addplot[name path=lower1, draw=none, forget plot] 
                table[header=false, col sep=comma, x index=0, y index=3] {data.txt};

            \addplot[NavyBlue, thick] 
                table[header=false, col sep=comma, x index=0, y index=1] {data.txt};
                \addlegendentry{MAP estimate}  
                
            \addplot[fill=NavyBlue!20, opacity=0.5, draw=none] 
                fill between[of=upper1 and lower1];
            \addlegendentry{95\% confidence interval}
            \end{axis}
    \end{tikzpicture}
    \caption{MAP subspace estimate and corresponding 95\% confidence interval as a function of the signal-to-noise ratio, using $N=500$ samples. The posterior converges to a Dirac measure at the true subspace as $\sigma_\epsilon \to 0$}
    \label{fig:bias}
\end{figure}

Assuming an uninformative uniform prior, we compute the posterior density over a subset of candidate subspaces as shown in Fig. \ref{fig:posterior_distribution} by sampling candidate subspaces \(\Gamma \in \operatorname{St}(2,1)\). Figure \ref{fig:angular_error_vs_N} illustrates that the angular error scales as we gather more is added. Figure \ref{fig:bias} demonstrates the evolution of the subspace angle MAP estimate as a function of the signal-to-noise ratio (SNR), $\mathrm{SNR} = y_0/\sigma_\varepsilon$. 

\subsubsection{Prediction} Next, we perform prediction as in Section~\ref{prediction}. We focus on the MAP conditioned distribution $p(y_f \mid w_a, \hat{\bf{S}})$ for computational feasibility as evaluating the full predictive posterior will require specialized sampling algorithms which we defer to an extension.

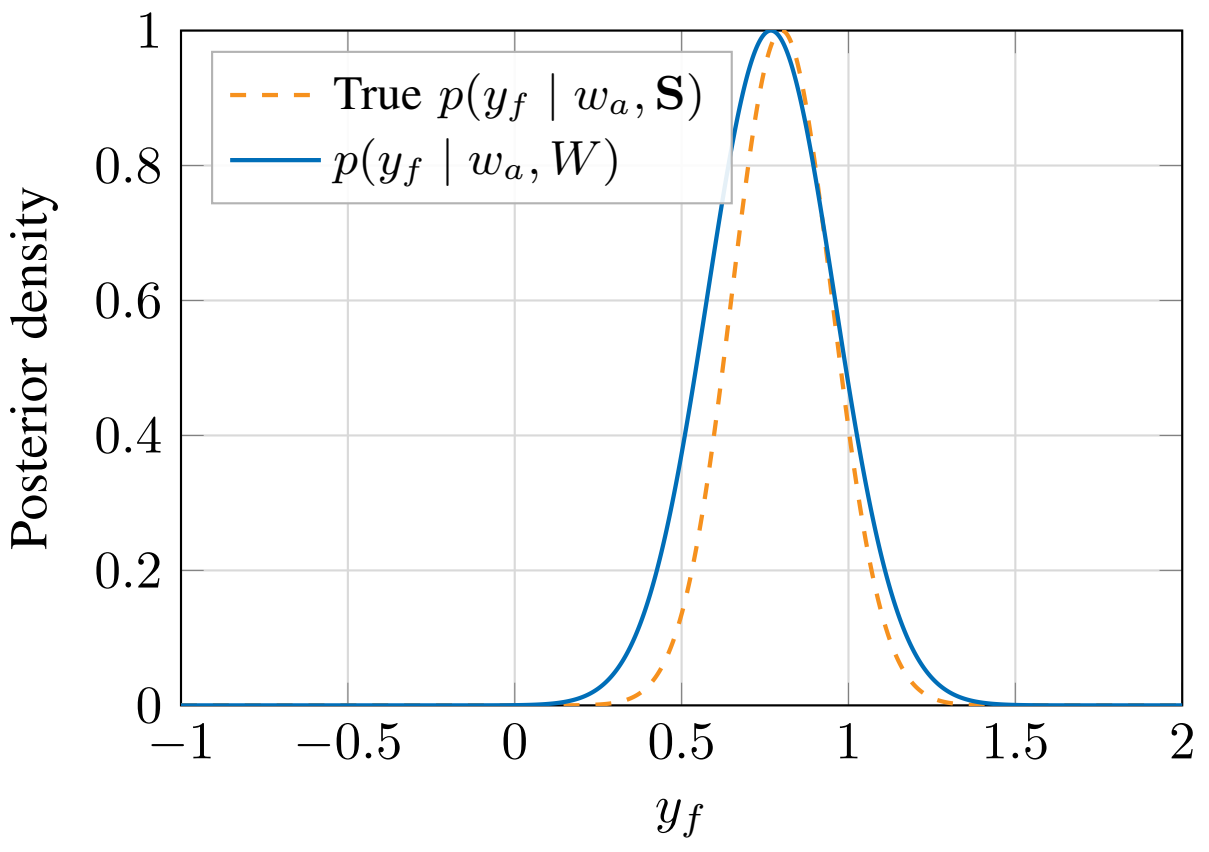
\begin{figure}[htbp]
  \centering
  \begin{tikzpicture}
    \begin{axis}[
      width=0.95\columnwidth,
      height=0.7\columnwidth,
      xmin=-1, xmax=2,
      ymin=0,    ymax=1,
      xlabel={$y_f$},
      ylabel={Posterior density},
      grid=major,
      grid style={gray!30},
      axis background/.style={fill=white},
      legend pos=north west,
      legend cell align={left},
      legend style={
        draw=gray!60,
        fill=white,
        fill opacity=0.9,
        text opacity=1,
        font=\small
      }
    ]
      \addplot[color=BurntOrange, thick, smooth, dashed] table[header=false, col sep=comma, x index=0, y index=1] {yf_posterior.txt};
      \addlegendentry{True $p(y_f\mid w_a, \mathbf{S})$}
      \addplot[color=NavyBlue, thick, smooth] table[header=false, col sep=comma, x index=0, y index=2] {yf_posterior.txt};
      \addlegendentry{$p(y_f\mid w_a, W)$}
    \end{axis}
  \end{tikzpicture}
  \caption{Posterior densities for $N=500$ with true noisy $y_f$ and posterior $y_f$ with $\sigma_\varepsilon=0.15$ and $\sigma_z=1$. using a fixed MAP subspace estimate rather than the full posterior distribution. The true and predicted distributions are close but offset, illustrating how errors in the estimated subspace propagate to the output prediction and introduce systematic bias. For $\sigma_\varepsilon=0.15$ and $y_0=1$, the posterior median of $y_1$ is 0.768 and its $95\%$ credible interval is [0.695, 0.840].}
  \label{fig:yf-posterior}
\end{figure}

\begin{figure}[htbp]
    \centering
    \begin{tikzpicture}
    \begin{axis}[
    width=0.95\columnwidth,
    height=0.7\columnwidth,
    xmode=log,
    xmin=5, xmax=100,
    xlabel={SNR $y_0/\sigma_\varepsilon$},
    ylabel={Output $y_\mathrm{f}$},
    grid = both,
    major grid style={gray!30},
    minor grid style ={gray!15},
      legend pos=south east,
      legend cell align={left},
      legend style={
        draw=gray!60,
        fill=white,
        fill opacity=0.9,
        text opacity=1,
        font=\small
      }
]
             \addplot[BurntOrange, dashed, thick, samples=2, domain=1:100] {0.8}; 
            \addlegendentry{True output mean}  
            \addplot[name path=upper_true, draw=none, forget plot] 
                table[header=false, col sep=comma, x index=0, y index=4] {MAP_vs_sigma_2.txt};
            \addplot[name path=lower_true, draw=none, forget plot] 
                table[header=false, col sep=comma, x index=0, y index=5] {MAP_vs_sigma_2.txt};
            \addplot[fill=BurntOrange!20, opacity=0.5, draw=none] 
                fill between[of=upper_true and lower_true];
            \addlegendentry{95\% confidence interval}
            \addplot[name path=upper1, draw=none, forget plot] 
                table[header=false, col sep=comma, x index=0, y index=2] {MAP_vs_sigma_2.txt};
            
            \addplot[name path=lower1, draw=none, forget plot] 
                table[header=false, col sep=comma, x index=0, y index=3] {MAP_vs_sigma_2.txt};

            \addplot[NavyBlue, thick] 
                table[header=false, col sep=comma, x index=0, y index=1] {MAP_vs_sigma_2.txt};
                \addlegendentry{MAP estimate}  
                
            \addplot[fill=NavyBlue!20, opacity=0.5, draw=none] 
                fill between[of=upper1 and lower1];
            \addlegendentry{95\% confidence interval}
            \end{axis}
    \end{tikzpicture}
    \caption{MAP output estimate and corresponding 95\% confidence interval as a function of signal-to-noise ratio, calculated using fixed MAP subspace estimate with $N=500$. This predictive distribution is conditioned on a single MAP subspace estimate and neglects uncertainty in the underlying subspace, producing overconfident predictions. The full predictive distribution \eqref{eq:predictive} should suffer less from this issue. However, the true prediction does reliably lie in the $95 \%$ confidence interval indicating an acceptable level of approximation error. }
    \label{fig:map_yf_vs_SNR}
\end{figure}
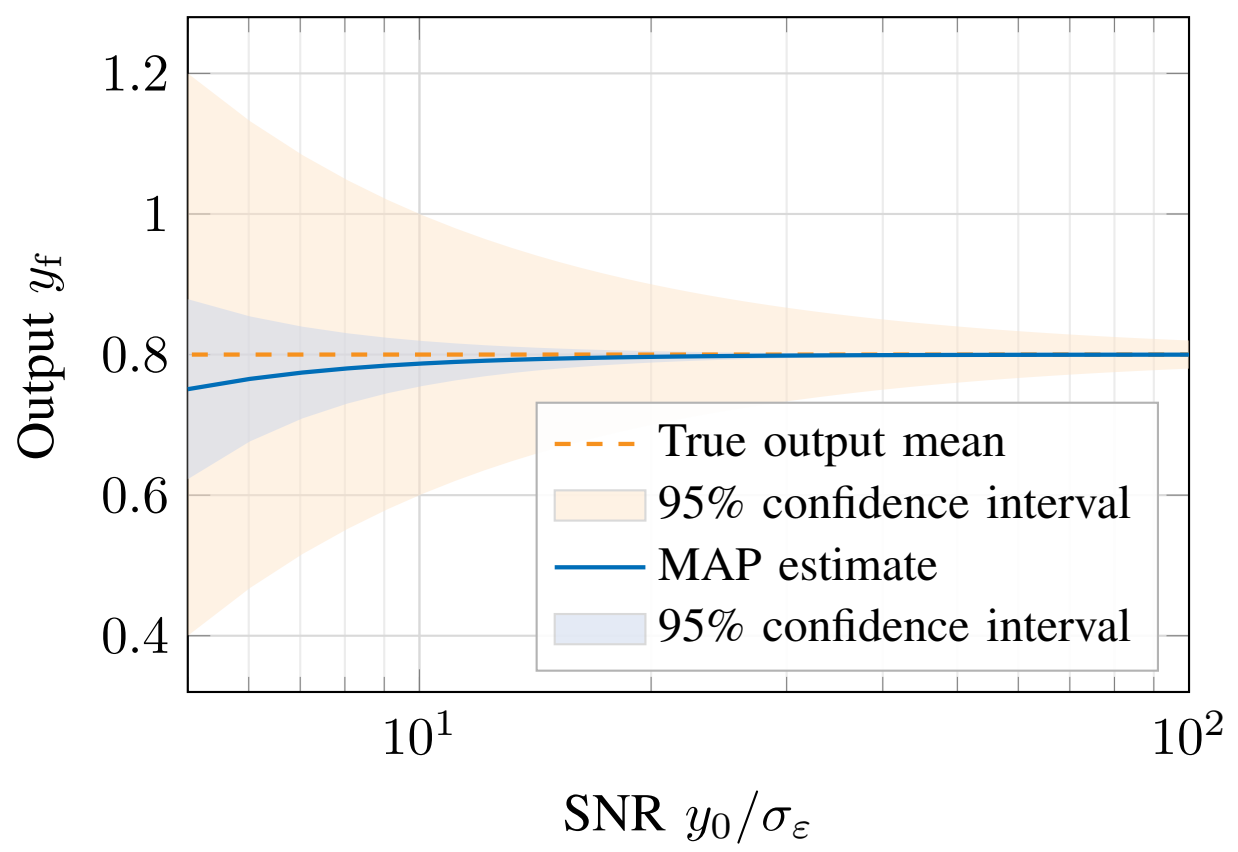

Here $d_a=r=1$ and, writing $\mathbf S=\operatorname{span}\{(\cos\theta,\sin\theta)\}$ with $\theta\in(-\pi/2,\pi/2]$, the noise-free predictor is $y_1=y_0\tan\theta$. Interestingly, the posterior density of $\theta$ is bounded below by a positive constant and $\int_{-\pi/2}^{\pi/2}|\tan\theta|\,d\theta=\infty$, so $y_1$ has no posterior mean. Since $\tan$ is increasing on $(-\pi/2,\pi/2)$, the posterior quantiles of $y_1$ are $y_0\tan$ of the posterior quantiles of $\theta$.  Figure \ref{fig:yf-posterior} compares the output distribution obtained using the true subspace \(\mathbf{S}\) with the predictive distribution obtained by propagating subspace uncertainty according to Theorem \ref{thm:posterior}. 
Finally, Figure \ref{fig:map_yf_vs_SNR} illustrates the dependence of the MAP-based output prediction on the SNR.

\subsubsection{Summary} At a high level, the numerical example indicates that: 1) the framework is consistent in the sense that the posterior distribution and predictive distribution approach that of the data generating process (almost surely) as $N\to \infty$ or $\sigma_\varepsilon \to 0$ and 2) the MAP conditioned distribution $p(y_f|w_a, \hat{\bf{S}})$ is overconfident but still provides a useful prediction and degree of uncertainty quantification.

\section{Conclusions and Future Work}\label{conclusion}
The paper introduced a Bayesian framework for learning restricted behaviours from noisy input--output data. Uncertainty about the restricted behaviour is represented by a probability distribution on the Grassmann manifold. Under the Gaussian observation model considered, we established conjugacy of the matrix Bingham family and derived predictive distributions that account for uncertainty in the inferred behaviour. A numerical case study illustrated the inference and prediction procedures.

Future work includes extending the observation model beyond isotropic noise, developing scalable algorithms for probabilistic prediction in larger systems, and exploring non-parametric representations of uncertainty about the restricted behaviour. Applications to model fusion and model predictive control could combine information from first-principles models with operational data while accounting for uncertainty in the inferred behaviour.

\bibliographystyle{IEEEtran}
\bibliography{myLibrary.bib}

\end{document}